\documentclass{iau}
\usepackage{graphicx}

\title[~~Compact nuclear objects - host galaxies relations] 
{Compact nuclear objects \\ and  properties of their parent galaxies}

\author {Anatoly V. Zasov$^1$ , Anatoly M. Cherepashchuk$^1$}   

\affiliation{$^1$ Sternberg Astronomical Institute, Faculty of Physics,
Lomonosov Moscow State University, University Prospect, 13, Moscow, Russia.}

\pubyear{2014}
\volume{304}  
\pagerange{}
\jname{Multiwavelength AGN Surveys and Studies}
\editors{}
\begin{document}

\maketitle

\begin{abstract}
 We consider the relationship between the masses of the compact nuclear objects in the centers of disky galaxies -- supermassive black holes (SMBHs) or nuclear star clusters (NCs) --  and such parameters  as the
 maximal velocity of  rotation
 $V_{\textrm{max}}$, obtained from the rotation curves, indicative dynamical mass $M_{25}$, and the  color 
index ($B{-}V$) of their parent galaxies.  It was found that the mass of nuclear clusters  $M_{\rm nc}$ correlates more 
closely with the velocity of rotation and total mass of galaxies  than the mass of supermassive black holes $M_{\rm bh}$. The dependence of masses of the central objects
on the color index is bimodal: galaxies of the red group 
(red-sequence), which have ($B{-}V) > 0.6{-}0.7$,  differ from bluer galaxies, 
by higher values of  $M_{\rm bh}$ for similar host-galaxy parameters. In contrast, in the diagrams for nuclear clusters the ``blue'' and 
``red'' galaxies form unified sequences.  It agrees with scenarios in 
which most red-group galaxies form as a result of
loss of interstellar gas in a stage of high nuclear activity in
galaxies whose central black-hole masses are high, exceeding $10^6 {-} 10^7 M_{\odot}$ 
(depending on the total mass of the galaxy). The active growth of nuclear star clusters possibly begun after the violent AGN activity.
\end{abstract}

\firstsection 
\section{Introduction}

There are two kinds of central massive objects observed at the dynamical 
centers of both disky and elliptical galaxies: supermassive black holes (SMBHs), and nuclear star clusters (NCs).  The genetic connection between NCs and SMBHs, their
formation mechanisms, and, especially, the properties of their evolution 
are poorly known, being a subject of active discussions. 

The observations show that both NC and SMBH masses are correlated with the
properties of the host galaxies (see, e.g., \cite[Graham et al. (2011)]{Graham11}, \cite[Ferrarese et al. (2006)]{Ferrarese06}, \cite[Zasov et al. (2011)]{Zasov11}, \cite[Zasov et al. (2013)]{Zasov13}, \cite[Scott \& Graham (2013)]{Scott13} and reference therein).  

In the current study, we have analyzed the connection between the masses
of nuclear star clusters and supermassive black holes with the 
kinematic parameters of their host 
disky galaxies, using the published rotation curves. Since the growth mechanisms for the NCs and SMBHs are directly or
indirectly connected with the evolution of stellar population, below we  
take into account the ($B-V$) color index, corrected for absorption according to \cite[Hyperleda database]{Hyperleda03}.
 We did not consider purely elliptical galaxies withouthe rotating disks here. 

Our samples of galaxies with known NC masses  $M_{\rm nc}$ and SMBH masses  $M_{\rm bh}$ are based on the list of \cite[Seth et al. (2008)]{Seth08} and  \cite[Graham (2008)]{Graham08} respectively. Both samples were supplemented by the data for some galaxies taken from the other papers. All the references and data sources may be found in our paper \cite[Zasov\&Cherepashchuk (2013)]{Zasov13}. Since the number of galaxies 
with $\log M_{\rm bh}$< 6 is very small, we added to the sample of SMBHs a few objects with 
low mass black holes, although for some of them only upper limits of $M_{\rm bh}$ are known. 

The maximal velocity of rotation $V_{\textrm{max}}$ of galaxies we consider was taken from the rotation curves presented by different authors  (see the references in \cite[Zasov\&Cherepashchuk (2013)]{Zasov13}). We define   $V_{\textrm{max}}$ as the velocity which corresponds to the plateau or
the maximum in the measured rotation curve, if it is located beyond the 2 kpc radius in order to  ignore  the local maxima of $V(R)$ caused by comact bulges. For several galaxies with low-mass SMBH without the rotation curves we took  $V_{\textrm{max}}$ obtained from the width of the HI line in \cite[Hyperleda database]{Hyperleda03}. To avoid the inflence of the asimmetric drift, we ignored those galaxies where 
the velocity dispersion outside the bulge is comparable with 
 $V_{\textrm{max}}$.

\section{Masses of central objects and the properties of host galaxies}

 In some cases  the black hole mass is measured (or the upper limit is found) in the same galaxy where the nuclear cluster is observed. It allows to compare directly the ratios of their masses $M_{\rm bh}$/$M_{\rm nc}$. This ratio is compared with the color index ($B{-}V$) of parent galaxies in Fig. 1.  It demonstrates  that in most cases  the mass of SMBH exceed the NC mass, although for ``red'' galaxies with relatively large color index we have the opposite. This difference does not imply that massive black holes grow faster in ``red'' galaxies, rather it reflects the known fact that the mass range for NCs is much narrower than for SMBHs. The most essential here is the kind of bi-modality: there are practically no massive black holes among the ``blue'' galaxies. May they be ``blue'' just because the mass $M_{\rm bh}$ is not high enough? 

It is essential that that in all diagrams we considered for $M_{\rm bh}$ (however not for $M_{nc}$!)  the overall sample of galaxies clearly splits into two groups separated by 
$(B{-}V)\approx 0.6-0.7$. Here we tentatively adopt $(B{-}V) = 0.65$ as the
boundary between ``red'' and ``blue'' galaxies. These color groups are well known as the red and blue sequences 
of galaxies in the universal bimodal distribution of color indices in the 
color--luminosity diagram (see, e.g.,~\cite[Balogh et al.(2004))]{Balogh04}. The blue group contains
galaxies with active star formation, whereas the red group 
is filled with passively evolving galaxies, in which star formation 
is very weak or absent. 

\begin{figure}[b]
\begin{center}
 \includegraphics[width=0.5\textwidth]{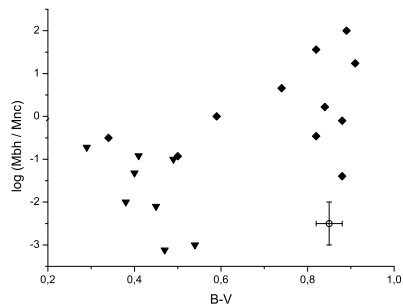} 
 \caption{Ratios $M_{\rm bh}$/$M_{nc}$ and color index of parent galaxies. Black triangles are  for galaxies with the upper limits of $M_{\rm bh}$. } 
\label{fig1}
\end{center}
\end{figure}

The dark halo could theoretically play a large 
role in SMBH formation, and in determining the final black hole mass.  The total (virial) mass of the dark halo is determined by the  circular velocity at the virial radius, which is close to  $V_{\textrm{max}}$.  So the relationship between the masses of the central objects, $M_{\rm bh}$ or $M_{\rm nc}$, and $V_{\textrm{max}}$  is of special interest.

The connection between the SMBH or NC masses and galactic circular
velocities has been considered in a number of studies, but the results are 
contradictory (see \cite[Kormendy et al. (2011)]{Kormendy11}). The tight correlation  $M_{\rm bh}$ -- $V_{\textrm{max}}$ for disk galaxies found in some studies  is evidently the result of using the
indirect estimates of $M_{\rm bh}$, based on 
their statistical dependencies on the central stellar-velocity dispersion (see the discussion in \cite [Zasov\&Cherepashchuk (2011))]{Zasov11}.  
 
Fig. 2  compares the masses of NCs (a) and SMBHs (b) with the velocity of rotation  $V_{\textrm{max}}$.  Dashed regression line in Fig 2b is the same as the solid line in Fig. 1a, being moved there for comparison. ``Red''  and ``blue'' galaxies are shown by the empty and filled symbols  respectively.

 The diagrams show that the masses of both NCs and SMBHs 
 correlate with the velocities of rotation, but the link between the mass and rotation is tighter for nuclear clusters than for black holes.  It is essential, that the ``red'' and ``blue'' galaxies are not separated in the NC plot, however they occupy distinct regions in the diagram for SMBHs: the 
``red'' galaxies (in Fig. 2b these are mostly lenticular galaxies) possess more massive black holes
than the ``blue'' ones with the same $V_{\textrm{max}}$. 

\begin{figure*}
\includegraphics[width=0.5\textwidth]{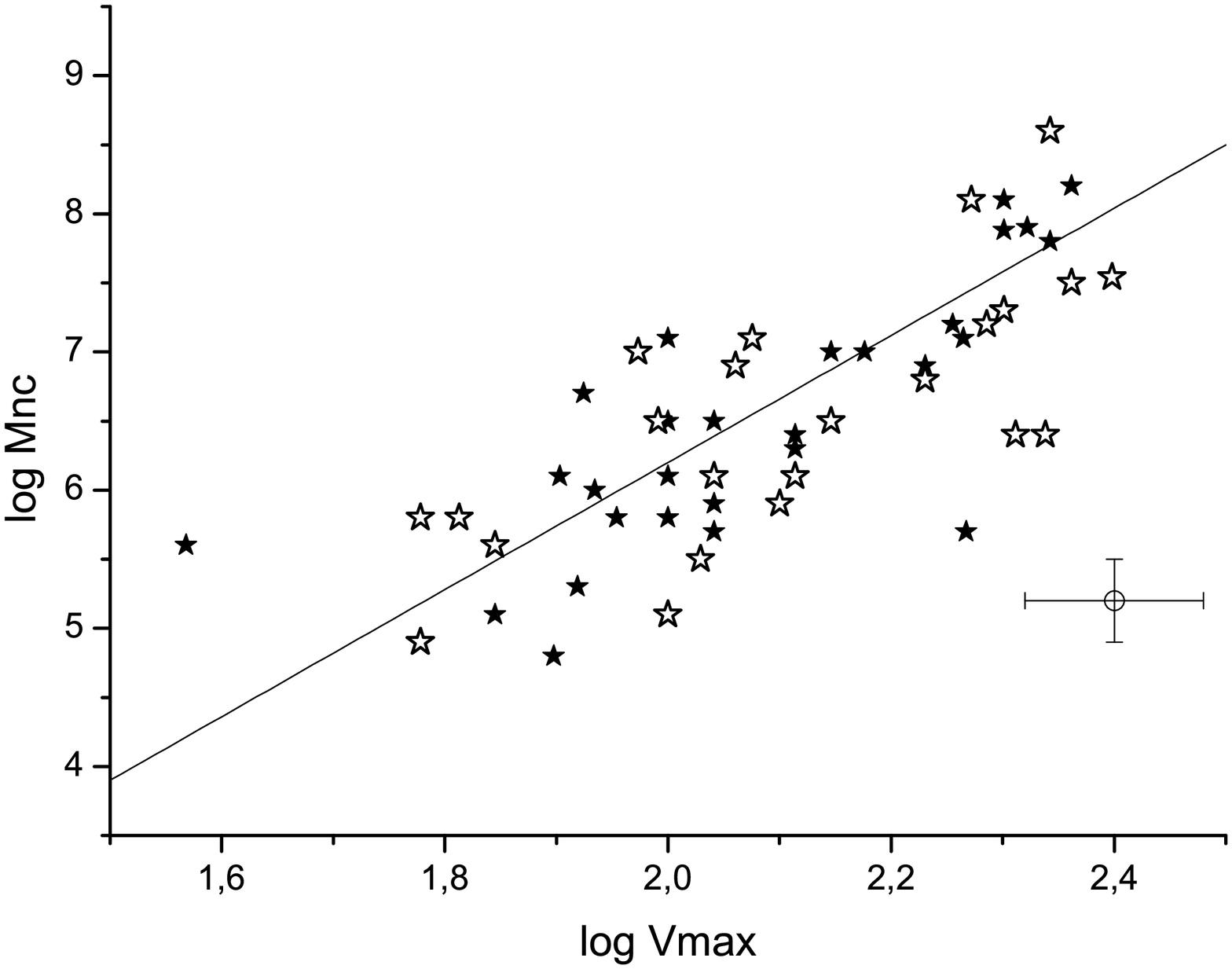}
\includegraphics[width=0.5\textwidth]{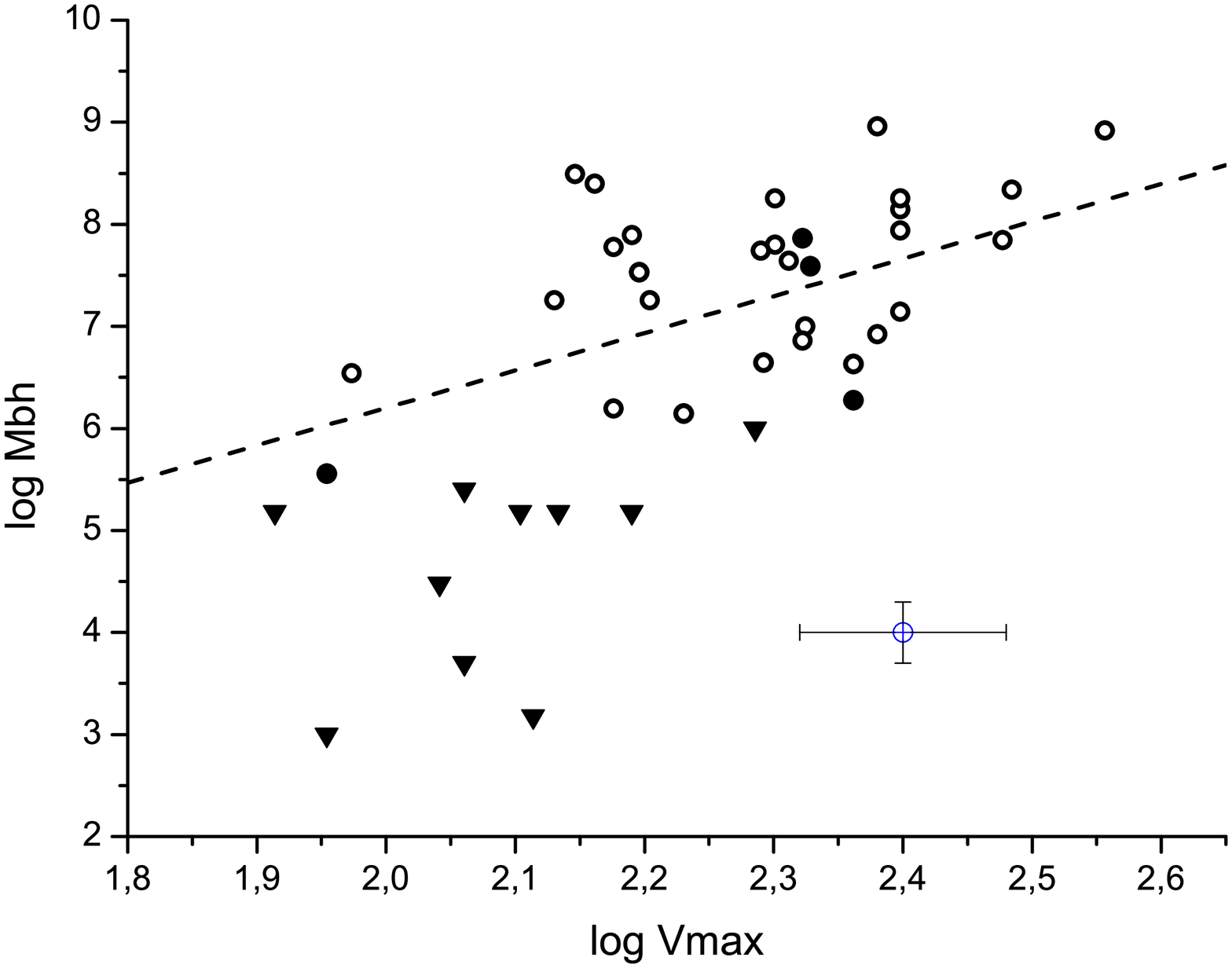}
\caption{ Dependence of (a) $M_{\rm nc}$ and (b) $M_{\rm bh}$ on the 
maximum (asymptotic) disk velocity. Filled triangles mark the galaxies with the upper limits of $M_{\rm bh}$. Regression line in Fig 2b is the same as in Fig. 2a, being moved there for comparison. ``Red''  and ``blue'' galaxies are shown by the empty and filled symbols  respectively. 	
}
\label{fig2}
\end{figure*}

Similar conclusions follow from a comparison of the mass of  central
objects with the total (indicative) mass of a galaxy within the optical
diameter $D_{25}$, defied as $M_{25} =
V_{\textrm{max}}^2D_{25}$/2G (Fig.~3).  Unlike $M_{\rm bh}$, the $M_{\rm nc}$ values for 
galaxies of both color groups form a unified sequence in all diagrams we considered. 

\begin{figure*}
\includegraphics[width=0.5\textwidth]{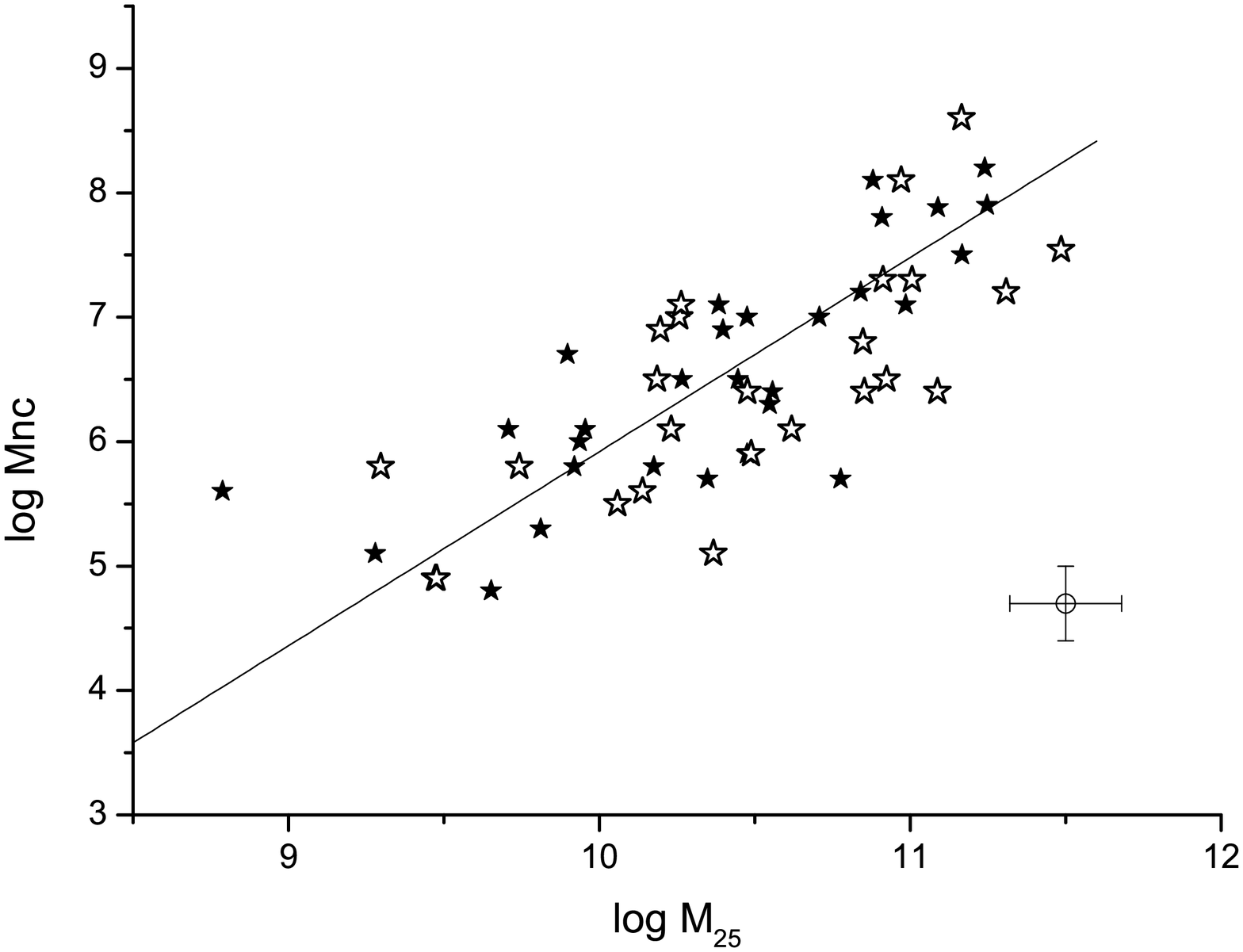}
\includegraphics[width=0.5\textwidth]{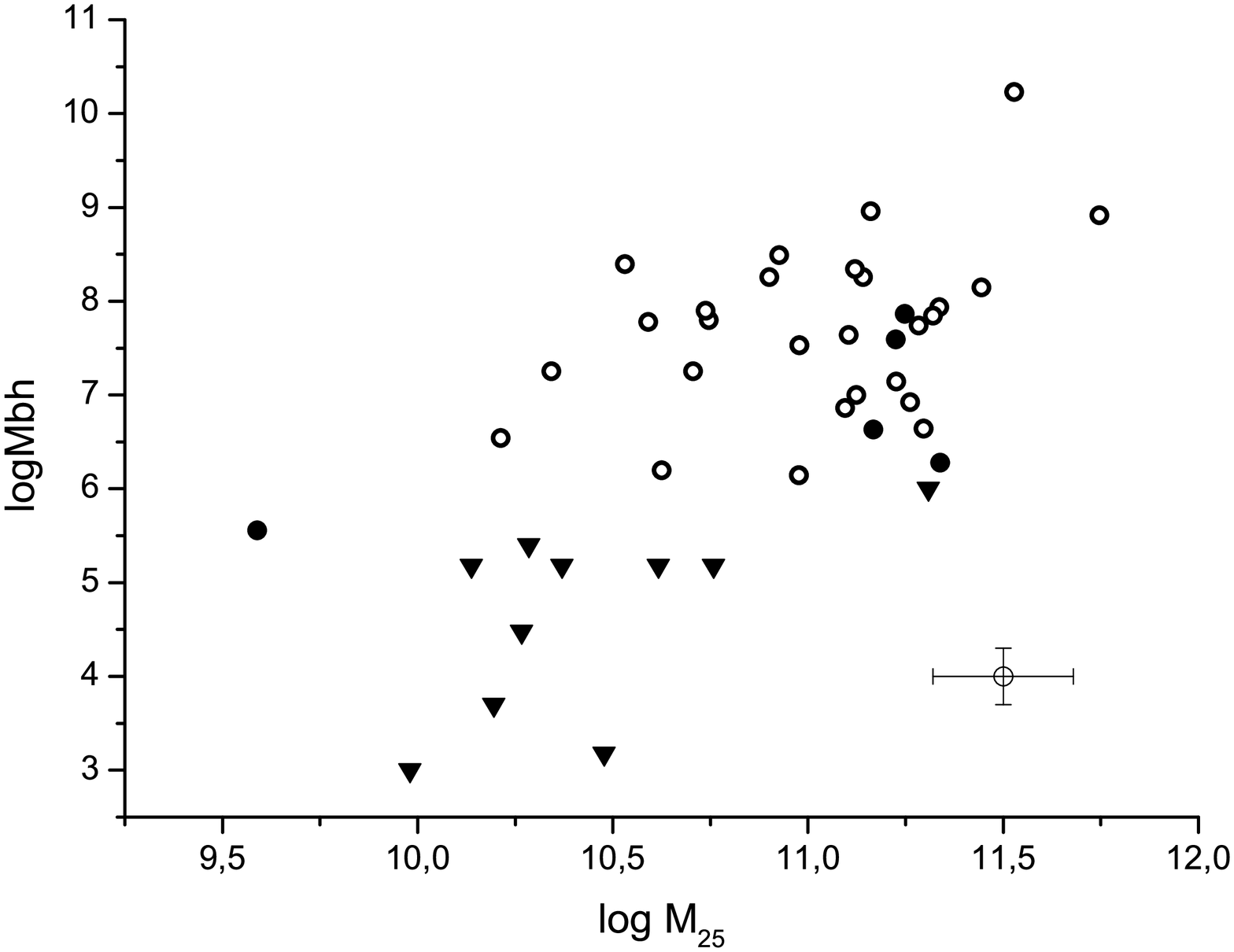}
\caption{Dependence of (a) $M_{\rm nc}$ and (b) $M_{\rm bh}$ on the
dynamic mass $M_{25}$ within the optical radius. The notation is
the same as in Figs. 2a,b.} 

\label{fig3}
\end{figure*}

\section{Discussion and conclusion}

The bimodal nature of color-index distribution of the observed galaxies indicates
that the  transition of galaxies from the blue to the red 
sequence must occur fairly quickly, at some stage of the evolution. 
Several hypothetical mechanisms for this transition have been considered 
(see the discussions of this problem by  \cite[di Matteo et al. (2005)]{Matteo05} and \cite[Mendez et al. (2011)]{Mendez11}).   The efficient mechanism that can halt
star formation in massive galaxies is the burst of AGN activity, which sweeps a gas out of the 
inner regions, which may be associated with galaxy mergers  (\cite[di Matteo et al. (2005)]{Matteo05},   \cite[Mendez et al. (2011)]{Mendez11} and \cite[Zubovas et al. (2012)]{Zubovas12}). 

It is important that the ``red'' and ``blue''  galaxies
are not distinguished on the diagrams where the NC masses are compared with the rotation velocity  $V_{\textrm{max}}$, or dynamic mass  of the host galaxies (see above), just as with  mass of stellar population, or with the velocity of rotation at a fixed distance from the nucleus (not shown here; see \cite[Zasov \& Cherepashchuk (2013)]{Zasov13}). In turn, correlations of  $M_{\rm bh}$ with the integral properties of patent galaxies have a different character: they are much looser and steeper than for  $M_{\rm nc}$. Moreover, SMBHs, unlike NCs, are usually more massive in the red-group galaxies, than 
in the blue-group galaxies, even if they have similar rotation velocities and masses. It agrees well with a scenario in which the decrease of  
star formation rate which makes a galaxy  ``red`` is the result of large mass of SMBH, which inspires a  high nuclear activity  at an early stage of the galaxy's evolution, and? as a result, moves a galaxy into the passively evolving red sequence.  Judging 
from the observed $M_{\rm bh}$ values for the ``red'' galaxies, the SMBH mass 
must exceed a few $\cdot 10^6 M_{\odot}$ to halt star formation, with a higher $M_{\rm bh}$ threshold 
required for more massive galaxies.

The impact of  active nucleus on 
the NCs is not entirely clear, however it is obvious
that the burst of activity may halt  the rapid growth phase of both SMBH and NC.  Since $M_{nc}$ for red-group and blue-group
galaxies correlate with the parent galaxy' properties in a similar way, it follows that the 
nuclear clusters of the ``red''  galaxies either managed to avoid the destruction during the active phase of nucleus, or (which is more likely)  they began their active growth after this violent event, which gave them possibility to evolve in a similar way in the ``red'' and ``blue'' galaxies.

\end{document}